\documentclass{article}

\usepackage{arxiv}

\usepackage[utf8]{inputenc} 
\usepackage[T1]{fontenc}    
\usepackage{hyperref}       
\usepackage{url}            
\usepackage{booktabs}       
\usepackage{amsfonts}       
\usepackage{nicefrac}       
\usepackage{microtype}      
\usepackage{lipsum}		
\usepackage{graphicx}
\usepackage{doi}
\usepackage[printonlyused]{acronym}  
\usepackage{hyperref} 
\usepackage[sort&compress,numbers]{natbib}
\usepackage{float}
\usepackage{graphicx}
\usepackage{amsmath}
\usepackage{mathrsfs}
\usepackage{amssymb}
\usepackage{cleveref}
\usepackage{multirow}
\usepackage{caption}
\usepackage{subcaption}
\usepackage{booktabs}
\usepackage{graphics}
\usepackage{enumitem} 
\usepackage{diagbox}
\usepackage{mdframed} 
\usepackage{tcolorbox} 
\usepackage{array}
\usepackage{mathtools}
\usepackage{bm}
\usepackage{adjustbox}
\newcolumntype{M}[1]{>{\centering\arraybackslash}m{#1}}
\usepackage{algorithm}
\usepackage{algpseudocode}

\title{Nonparametric Multivariate Profile Monitoring Via Tree Ensembles}

\author{ \href{https://orcid.org/0000-0001-8755-5233}{\includegraphics[scale=0.06]{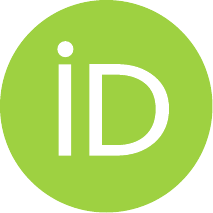}\hspace{1mm}Daniel A.~Timme} \\ 
	Department of Statistics\\
	Florida State University\\
	Tallahassee, FL 32308 \\
	\texttt{dat20bu@fsu.edu} \\
	\And
	\href{https://orcid.org/0000-0001-8196-7229}{\includegraphics[scale=0.06]{orcid.pdf}\hspace{1mm}Andr\'{e}s F.~Barrientos} \\
	Department of Statistics\\
	Florida State University\\
	Tallahassee, FL 32308 \\
	\texttt{abarrientos@fsu.edu} \\
	\And
	\href{https://orcid.org/0000-0003-3977-3601}{\includegraphics[scale=0.06]{orcid.pdf}\hspace{1mm}Eric Chicken} \\
	Department of Statistics\\
	Florida State University\\
	Tallahassee, FL 32308 \\
	\texttt{chicken@fsu.edu} \\
	\And
	\href{https://orcid.org/0000-0000-0000-0000}{\includegraphics[scale=0.06]{orcid.pdf}\hspace{1mm}Debajyoti Sinha} \\
	Department of Statistics\\
	Florida State University\\
	Tallahassee, FL 32308 \\
	\texttt{dsinha@fsu.edu} \\
}

\renewcommand{\shorttitle}{Timme et al.}

\hypersetup{
pdftitle={A template for the arxiv style},
pdfsubject={q-bio.NC, q-bio.QM},
pdfauthor={David S.~Hippocampus, Elias D.~Striatum},
pdfkeywords={Profile Monitoring, Decision Trees, Statistical Process Control, Kolmogorov-Smirnov, Nonparametric Model, Multivariate},
}

\begin{document}
\maketitle

\begin{abstract}
	Monitoring random profiles over time is used to assess whether the system of interest, generating the  profiles, is operating under desired conditions at any time-point. 
	In practice, accurate detection of a change-point 
	within a sequence of responses that exhibit a functional 
 relationship with multiple explanatory variables is an important goal for effectively monitoring such profiles. 
 We present a nonparametric method utilizing ensembles of regression trees and random forests to model
 the functional relationship along with associated Kolmogorov-Smirnov statistic to monitor profile behavior. 
Through a simulation study considering multiple factors, we demonstrate that our method offers strong performance and competitive detection capability when compared to existing methods.
\end{abstract}


\section{Introduction}
Statistical Process Control (SPC) deals with sequentially monitoring a process to verify whether the underlying system, such as the production line or traffic or service facility, is stable and functioning properly over time\cite{SPC}. 
Modern manufacturing, surveillance, and service systems are frequently characterized by high complexity. Consequently, the task of monitoring the process for detecting errors, deterioration, or anomalies becomes challenging, particularly in this era of automated and high-dimensional data collection.
In this context, the monitoring process over time is performed via detecting structural changes in multiple characteristics of the system's quality or state. 
For more background on profile monitoring, we suggest referring to the reviews available in  Qiu \cite{SPC}; Noorossana, Saghaei, and Amiri \cite{Nooro}; Woodall, Spitzner, Montgomery, and Gupta \cite{Wood-Sptiz-Monty}; Woodall \cite{Woody_07}; Woodall and Montgomery \cite{Woody-Monty}.

There are numerous applications in practice of profile monitoring within SPC, including semiconductor manufacturing \cite{gardner97}, automobile engine testing \cite{amiri_09}, stamping operations \cite{jin_99}, curvature evaluation of a mechanical component \cite{colosimo_08}, supply chain, assembly processes, aeronautics, banking systems, health care, and the internet \cite{SPC}. Florac, Carlton, and Bernard \cite{Florac_00} used SPC to monitor various relevant processes as part of the Space Shuttle Onboard Software Project. 
Non-physical systems can also be monitored via SPC. For example,  cyber-threats can degrade both physical and non-physical systems \cite{Cab1}. 
Ultimately, SPC has multiple applications for online monitoring of both physical and non-physical systems. The accurate identification of a malfunction can save time and reduce overhead costs.

A system operating only under natural variability ("background noise") \cite{Monty_09} is said to be statistically in-control. 
When the variability beyond background noise, typically caused by external factors, 
leads to the system performing in an unacceptable manner, 
 it is considered statistically out-of-control  \cite{SPC, Nooro}. 
A monitoring statistic is a summary of the observed profiles and explanatory variables. When a process is statistically in-control, the monitoring statistics 
should lie within an Upper Control Limit (UCL) and a Lower Control Limit (LCL). 
These two values are determined during initialization and usually using some set of historical profiles \cite{Nooro}. 

In this paper, our focus is on developing an SPC procedure specifically designed for situations where the quality characteristic of the underlying process, at each monitoring time $t$, is represented by a functional relationship between a response variable and multiple explanatory variables.
Due to their flexibility and broad applicability in practical scenarios, we employ non-parametric tools in our approach. Specifically, our proposed monitoring method combines the Kolmogorov-Smirnov (KS) statistic to compare residual distributions across different time points and regression-trees-based approaches to model,  at each time point, the aforementioned functional relationship. The monitored residuals are computed as the difference between the observed response and the estimate of the functional relationship at each time point.
Most of the existing monitoring methods use parametric methods, which provide good results when the key parametric assumptions are correct. However, 
verifying the validity of these assumptions is generally challenging in practical settings,
especially for applications with frequently monitored processes.
The benefit of our non-parametric monitoring method is that it does not make restrictive modeling assumptions about the underlying distribution of the profiles. 
Additionally, we find that while there is currently a plethora of literature in profile monitoring with a single explanatory variable \cite{WilliWoodyBirch, ChickPigSimpson, Shing-Yada, NikooNooro, McChicken, ChickHillPig, varbanov2019bayesian, MagnificoGrass, ChuangHungYang, HadSamHamid, ZouQiuHawk, YangZouWang, QieZou}, research on profile monitoring with multiple explanatory variables is relatively scarce \cite{HungTsaiYangChuang, Lietal, Iguchi}. It is worth noting that our proposed method can effectively handle multiple explanatory variables, addressing this research gap.

In the \autoref{Sec:Background}, we provide a brief review of the existing methods and state which we compare our methodology against. 
Subsequently, we present regression-trees-based models and the KS statistic, which we utilize to design our proposed monitoring methodology in \autoref{sec:method}. In \autoref{sec:method}, we also describe the strategy employed to determine the control limit. Our simulation study in \autoref{sec:sim} compares the proposed method to other existing methods in the literature. The simulation setup is outlined in \autoref{subsec:simsetup}, and the results of our simulation study and comparisons are presented in \autoref{subsec:simresults}.
We conclude with a discussion of our findings and identify the limitations of our method in \autoref{sec:discuss}.

\section{Background}\label{Sec:Background}

In this paper, the explanatory variables $\bm{x}_{i}^{t}$ and responses $y_{i}^{t}$ are sequentially observed in $t$. The responses make up the profiles which are used for making decisions about the status of the process at discrete monitoring time $t$. Specifically, at each discrete monitoring time $t$, we assume that we observe profiles of the form
\begin{align}\label{eq:profile}
    y_{i}^{t} = f^{t}(\bm{x}^{t}_{i}) + \varepsilon_{i}^{t},  
\end{align}
\noindent with the functional relationship $f^{t}: \mathbb{R}^{p} \rightarrow \mathbb{R}$, where $\bm{x}_{i}^{t} \in \mathbb{R}^{p}$ for $ i=1,2,\cdots ,n$ are  independently sampled multi-dimensional explanatory variables, $\varepsilon_{i}^{t} \in \mathbb{R}$ is the random noise associated with the process, and $y_{i}^{t} \in \mathbb{R}$ is the response corresponding to the vector $x_{i}^{t}$ of the explanatory variables. 
We denote the vector of $n$ observed responses and the matrix corresponding $n$ vectors of explanatory variables  
in \autoref{eq:profile} at time $t$ as 
$\bm{y}^{t}$
and $\bm{x}^{t}$, respectively.

The overall monitoring algorithm is broken up into the retrospective and prospective phases, or Phase I and II \cite{Nooro}. In Phase I, the retrospective step, a practitioner analyzes a set of historical profiles to ensure they are in-control. In Phase II, the sequential process monitoring uses data at each monitoring time to identify if the process is in-control or out-of-control. If the monitoring statistic is within the interval of LCL and UCL, the process is believed to be in-control; otherwise, it is out-of-control. When the process becomes truly out-of-control, the goal is to promptly identify this change to prevent the continued production of, say, a lower quality product than desired. 
Our proposed approach primarily focuses on Phase II and centers around the development of nonparametric profile monitoring methods.

There are some existing works in this area with various models and monitoring statistics (see \cite{WilliWoodyBirch}-\cite{QieZou} and references therein). 
However, most of these works deal with a scalar (or univariate) explanatory variable, while there does not appear to be a lot of work with multivariate profile monitoring. 
Williams et al. define three statistics using fitted responses averaged over a set of $m$ historical profiles \cite{WilliWoodyBirch}. Several articles have used wavelets for the regression function with a univariate/scalar explanatory variable, as in Chicken et al.  \cite{ChickPigSimpson} who monitor the $\ell_{2}$-norm of the difference between wavelet coefficients of an observed and in-control profile. Chang and Yadama in \cite{Shing-Yada} denoise observed profiles by applying a discrete wavelet transformation with thresholding and model said profile with a B-spline. Nikoo and Noorossana in \cite{NikooNooro} monitored the profile mean using  Hotelling $T^{2}$ statistics of wavelet coefficients and the profile variance using median absolute deviation of the highest level-detail. McGinnity et al. in \cite{McChicken} used a weighted sum of averaged wavelet coefficients before and after change-point. 
See \cite{varbanov2019bayesian} for a Bayesian monitoring approach using the posterior distribution of change-point location of the wavelet coefficients.
For clustered responses, Chicken et al. in \cite{ChickHillPig} apply the $k$-means clustering algorithm to the coefficients. 
As examples of other profile monitoring methods, Grasso et al. monitor the sum of prediction errors and Hotelling $T^{2}$ on warping coefficients and model profiles using functional principal component analysis scores \cite{MagnificoGrass}. Chuang et al. use bootstrapping on in-control data to build a confidence region and observe if profiles region fitted via B-splines fall outside of said confidence region \cite{ChuangHungYang}. 
Hadidoust et al. modeled profiles with a smoothing spline and applied a Hotelling $T^{2}$ statistic to the coefficients to monitor the profiles \cite{ HadSamHamid}. Zou et al. used a local linear kernel estimator to model nonlinear profiles and monitor the function with a generalized likelihood ratio statistic \cite{ZouQiuHawk}. Yang et al. monitor the exponentially weighted moving average (EWMA) on standardized residuals \cite{YangZouWang}. Unlike the methods mentioned above, Yang et al. use a dynamic UCL. Qiu and Zou also consider a EWMA approach, though theirs is based on local linear kernel smoothing to monitor profiles \cite{QieZou}. 
These mentioned above are only some of the methods during recent years 
focusing on scalar predictor. 

Compared to the literature on scalar predictor, the works on multivariate predictors are far more sparse. Hung et al.'s \cite{HungTsaiYangChuang} method uses support-vector regression (SVR) to model the profiles with multiple predictors and a moving block bootstrap region to monitor. Li et al.'s method \cite{Lietal} also uses an SVR to model profiles, though they use the nonparametric statistics of Williams et al. \cite{WilliWoodyBirch} and an EWMA control chart for monitoring.
%
In a more recent paper \cite{Iguchi}, Iguchi et al. used the $\ell_{2}$-norm of index coefficients of Single Index Model (SIM) of the regression function to obtain better  detecting of an out-of-control process than the methods employed by Li et al. Additionally, their SIM based method's false alarm rate (FAR) is comparable with the methods presented by Li et al. 
We will compare our method with the statistics used by Li et al. \cite{Lietal} 
and the SIM-based methods proposed by Iguchi et al. \cite{Iguchi}.

\section{Methodology}
\label{sec:method}

In this section, we describe our proposed approach. First, in \autoref{subsec:changepoint}, we present the hypothesis that we want to test for detecting change points, along with some statistics of relevant importance in SPC. These statistics help assess the performance of any approach and facilitate comparisons with existing ones. Then, in \autoref{subsec:modelingdata}, we explain how we use regression trees and random forests to estimate the functional relation $f^t$ of \autoref{eq:profile}, which is subsequently used to compute residuals. The monitoring statistic we use to monitor the residual distribution across time is described in \autoref{subsec:monitoringstatistic}.


\subsection{Change-Point Detection}\label{subsec:changepoint}
The change-point detection method is commonly used in conjunction with SPC. Here, the goal is to correctly identify whether a system is currently out-of-control at time $T$ based on observed profiles, which also include a given set of $m$ in-control historical profiles. In other words, we aim to test the following hypothesis:
\begin{align}\label{eq:ChgPtHyp}
    \begin{split}
        H_{0}: f = f^{1} = \cdots =& f^{T} \\
        H_{A}: f = f^{1} = \cdots =& f^{\tau} \ne f^{\tau +1} = \cdots = f^{T} 
        \ \text{for some}\  0\leq \tau<T\ ,
    \end{split}
\end{align}
where $\tau\ge0$ is the unknown last in-control time-step. 
For the functional relationship $f^t$ in \autoref{eq:profile}, we now make the assumption that $f^t = f$ when the process is in-control (i.e., $t \leq \tau$) and $f^t = \phi$ when the process is out-of-control (i.e., $t > \tau$). Therefore, if $\tau$ is the last time the process was in-control, we assume that the data-generating mechanism corresponds to
\begin{align}\label{eq:icooc}
   y_{i}^{t} = 
        \begin{cases}
            f(\bm{x}_{i}^{t}) + \varepsilon_{i}^{t} &\text{for}\ t \le \tau \\
            \phi(\bm{x}_{i}^{t}) + \varepsilon_{i}^{t} &\text{for}\  t > \tau \ .
        \end{cases}
\end{align}
To decide whether or not to reject the null hypothesis of \autoref{eq:ChgPtHyp} for the change-point framework at sequentially each time-step $T$, we will perform our methodology based on the comparison of the distributions 
of two sets of residuals--one set of residuals from a model fitted on (presumably) in-control data and the other set from the function fitted to the currently observed profile. This method is explained later in detail.

\begin{figure}[t]
  \centering
  \includegraphics[width=0.86\textwidth]{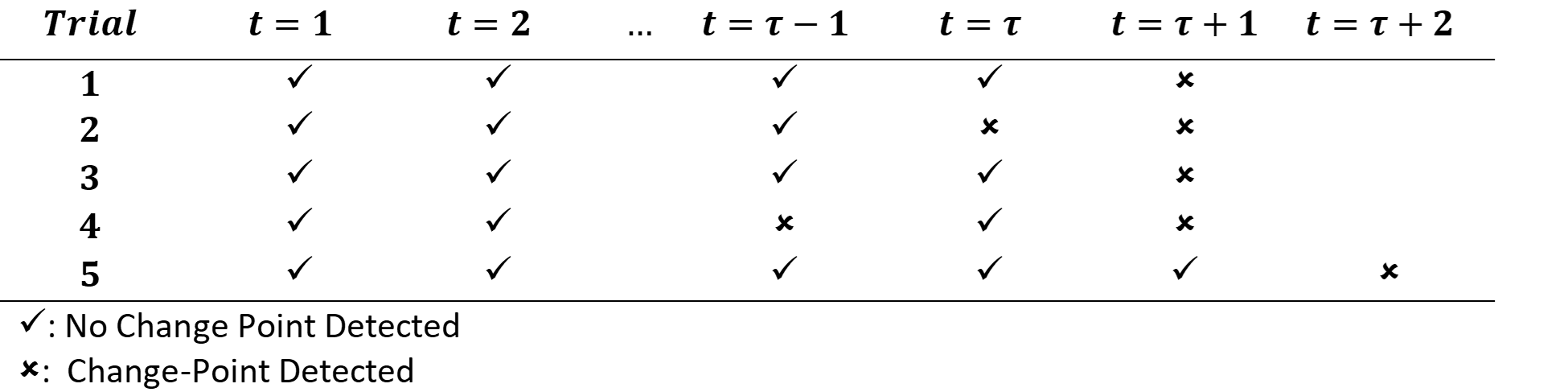}
  \caption{Change-point detection toy example}
  \label{fig:CPDToyEx}
\end{figure}

When a process is out-of-control, we ideally wish to reject the null hypothesis of \autoref{eq:ChgPtHyp} when $T$ is equal to $\tau+1$ to avoid continued manufacturing of a product which is of lower quality than desired. 
The performance of a change-point monitoring method can be evaluated simply by taking the Monte Carlo (MC) approximation of ARL$_1$ given by   
\begin{equation}\label{eq:ARL1}
    \text{ARL}_{1} \simeq \frac{1}{N} \sum_{j=1}^{N} \left( T_{j}-\tau \right)
\end{equation}
using $N$ idependent trials, where $T_{j}>\tau$ is the run length in the $j$-th trial until the first correct acceptance of $H_A$ when the true process is out-of-control. 
As a simple illustration, for the toy example in \autoref{fig:CPDToyEx}, 
we see that $T_{j}=\tau + 1$ for $j=1,2,3,4$ and $T_{5}=\tau + 2$ for $N=5$ trials, resulting in the MC approximation of \autoref{eq:ARL1} as $(1+1+1+1+2)/5=1.2$.

When an in-control process is  incorrectly flagged 
by the monitoring method as out-of-control, it is referred to as a false alarm (FA). 
Similar to Type 1 error in hypothesis testing, any false alarm is not desirable because it can lead to unnecessary delay due to stopping the in-control system. 
The False-Alarm Rate (FAR), the probability of 
a false alarm at a monitoring time $t<\tau$,  is another useful performance measure with the MC approximation 
\begin{equation}\label{eq:FArate}
    \text{FAR} \simeq \frac{N_{FA}}{N + N_{FA}}\ ,
\end{equation}
where $N_{FA}=\sum_{j=1}^N{R_{j}}$ with $R_{j}$ being the number of times the monitoring process has been restarted for incorrectly flagging the process 
as out-of-control before time $\tau+1$ during $j$-th 
trial.
At each of the $N$ independent trials used for MC approximation of FAR in \autoref{eq:FArate}, 
after each incorrect flagging of the process as out-of-control at any time before $\tau+1$,
the process monitoring is restarted (this explained in more detail in an upcoming section).
For the toy example in \autoref{fig:CPDToyEx} 
with $N=5$, if there are not false alarms in any of 5 
trials, then \autoref{eq:FArate} will be $0/(5+0)=0$. However, in this example we see that 
incorrect flagging of change-point 
occurred at $t\leq \tau$ in trials $2$ and $4$, 
with no further incorrect flagging before $\tau+1$ 
after restarting the monitoring.
That is, there are $N_{FA}=2$ false alarms and thus $2$ restarts with \autoref{eq:FArate} being $2/(5+2)\approx0.2857$.

\subsection{Modeling the Data} \label{subsec:modelingdata}

\begin{figure}
  \centering
  \includegraphics[width=0.85\textwidth]{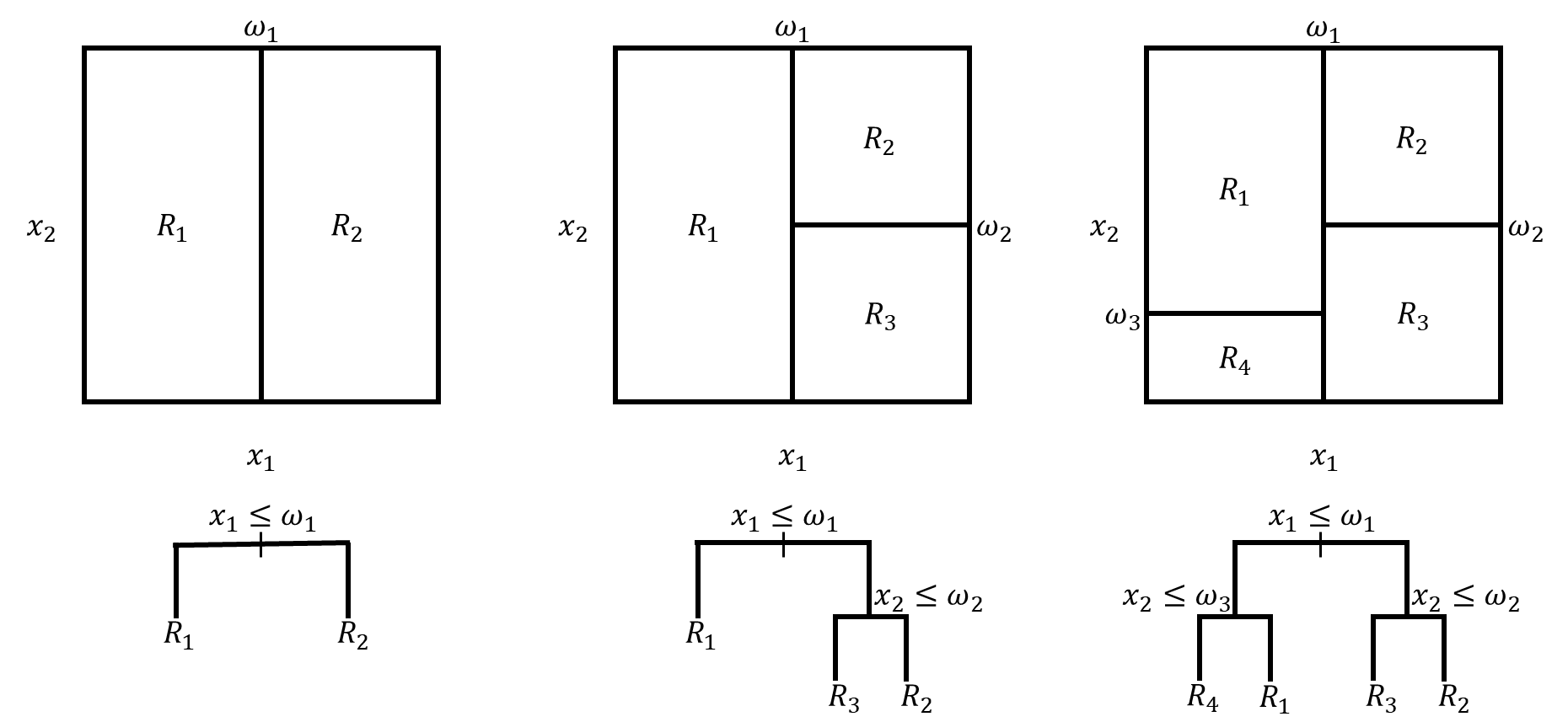}
  \caption{Generic regression tree example with regions, $R_k$}
  \label{fig:RTEx}
\end{figure}

Regression trees are commonly used prediction tools in machine learning, which are typically applied 
for continuous response variables. 
They are widely employed in modeling due to their ability to provide accurate predictions while being straightforward to construct, comprehend, and interpret. Suppose we have $i=1,\cdots,n$ observations of the explanatory variables $\bm{x}_i^t$ and corresponding profiles $y_i^t$. Additionally, suppose we have a partition $R_{1}^{t},R_{2}^{t},\cdots,R_{K}^{t}$ of the space of explanatory variables into $K$ regions. Using a minimization of the sum of squares, $c_{k}^{t}$ is estimated as the average of the profiles, ${y}_i^t$. Now, we can model the response as piece-wise constant $c_{k}^{t}$ within region $R_{k,t}$ as 
\begin{equation}\label{eq:RegTreef}
     \mathcal{T}^t(\bm{x}_i^t) = \sum_{k=1}^{K} c_{k}^{t} I_{  R_{k}^{t} } \left(\bm{x}_i^t \right) ,
\end{equation}
where $I_{R_{k}^{t}}(\bm{x}_{i}^{t})$ is an indicator function of $\bm{x}_{i}^{t} \in R_{k}^{t}$. \autoref{fig:RTEx} shows three generic examples of a regression tree for $2$ independent variables. Here, the partitions $R_1^t,\cdots,R_K^t$ constitute the structure/topology of the $t$-th decision tree. The observations which satisfy the condition at each decision point is assigned to the branch on the left. 


{\color{black} Random forests are another powerful machine-learning technique that make predictions using an ensemble of regression trees \cite{breiman2001random}. They combine the ensemble outputs by averaging to provide robust and reliable results. Their advantage over individual regression trees lies in their ability to reduce overfitting and improve predictive accuracy. For the sake of brevity, we do not provide a detailed description of how random forests are constructed, but we refer the reader to \cite{breiman2001random,gareth2013introduction} for a comprehensive explanation. Henceforth, and with some abuse of notation, we will use $\mathcal{T}^t(\bm{x}_i^t)$ to represent the prediction at time $t$ based either a regression tree or a random forest at $\bm{x}_i^t$.}

Assuming that the process has been in-control until time $t-1$, we estimate $f$ at time $t$ by the following regression tree (or random forest) ensemble, that is,
$$
\hat{f}^t\left(\bm{x}_{i}^{t}\right)=\frac{1}{t-1+m} \sum_{j=-m+1}^{t-1} \mathcal{T}^{j} \left(\bm{x}_{i}^{t}\right).
$$
Notice that this average is over $(t-1) + m$ regression trees (or random forests), which are fitted on $m$ in-control historical profiles, plus the $(t-1)$ previous profiles that have been observed since process monitoring began. We use $\hat{f}^t\left(\bm{x}_{i}^{t}\right)$ as a prediction of $\hat{y}_{i}^{t}$ and compute the residuals using $\hat{e}_{i}^{t}=y_{i}^{t}-\hat{y_{i}}^{t}$. The premise here is that as long as the process is in-control, the distribution of the residuals should remain the same over time. However, if the process goes out-of-control at time $t$, the distribution of the residuals is expected to change at that time. For this reason, we use a monitoring statistic that allows us to assess whether there have been changes in the distribution of the residuals at a given time. The next subsection describes in detail how we define our monitoring statistic, which is motivated by the KS test statistic.


\subsection{Selected Monitoring Statistic}\label{subsec:monitoringstatistic}

To monitor changes in the distribution of residuals over time, we begin by examining a statistic that quantifies the difference between estimates of the residual distributions at time $t$ and time $j$, where $j < t$. The residuals are computed according to the method outlined in \autoref{subsec:modelingdata}. To estimate such distributions, we use the empirical cumulative distribution function (CDF) of the residuals at each time, denoted as $\hat{F}_e^t$ and $\hat{F}_e^j$. The distance is computed as
\begin{equation}\label{eq:KSstatistics}
D\left(\hat{F}_e^t, \hat{F}_e^j\right) = \sup_{z \in \mathbb{R}} \lvert\hat{F}_e^t(z) - \hat{F}_e^j(z)\rvert,
\end{equation}
which is also employed to define the well-known Kolmogorov-Smirnov test \cite{Kolmogorov, Smirnov}. We refer to $D\left(\hat{F}_e^t, \hat{F}_e^j\right)$ as the KS statistic. Notice that if the process has been in-control until time $t$, then we expect the distances between $\hat{F}_e^t$ and $\hat{F}_e^j$, for $j = -m+1, \ldots, t-1$, to be ``small." However, if those distances suddenly increase at time $t$, it provides evidence that the process has experienced a change, which likely indicates it is out-of-control. Motivated by these ideas, we propose using the following monitoring statistic,
\begin{equation}\label{eq:monitoring_statistic}
\xi^{t}=\max\left\{D\left(\hat{F}_{e}^{t},\hat{F}_{e}^{j}\right) \lvert j=m-1,\cdots,t-1 \right\}.
\end{equation}
The idea behind using the maximum of KS statistics, $\xi^{t}$, as a monitoring statistic is to capture the most significant change in the distribution of residuals over earlier times. If this change remains ``small,'' then the process is assumed to be in control; otherwise, it is likely to be out-of-control. 
The steps necessary to compute the proposed monitoring statistic $\xi_t$ are summarized in \autoref{alg:Meth}.
Another important step in process monitoring is identification of the control limit. An Upper Control Limit (UCL) is calibrated to some desired Average Run Length, ARL$_0$ under in-control process. In practice, a value of ARL$_{0}\approx200$ is often used. At each time-step of SPC, we are interested in creating a 
decision rule with two possible decisions: alarm for out-of-control profile when our monitoring statistic is greater than the UCL, otherwise, no alarm. 
If the data is generated by an in-control distribution, 
we are interested in the probability of a false alarm (in this case, same as the probability of misclassification). 
Since we are using the maximum of KS statistics as a monitoring statistic (see \autoref{eq:monitoring_statistic}), one might consider employing a union bound to determine an upper bound for the UCL. Such a bound can be derived using the union bound with the Glivenko-Cantelli theorem \cite{Glivenko, Cantelli} or a Chernoff-type bound \cite{Chernoff}.
Unfortunately, the validity of these theoretical bounds requires that the KS statistics be independent across time. Because the computation of the KS statistics defining \autoref{eq:monitoring_statistic} always involves $\hat{F}_e^t$, the independence assumption is not valid. Another aspect that creates dependence in the KS statistics across time is the fact that the residuals are computed using regression trees or random forests fitted at previous times. For this reason, we employ bootstrap-based simulations to estimate the UCL, which allows us to account for such dependence. Leveraging our assumption of having $m$ historical in-control profiles, we generate simulated profiles by sampling data points from these $m$ historical profiles. The procedure, summarized in \autoref{alg:BS}, calculates the run length for a specified control limit. Through the use of this algorithm and a grid search, we determine the corresponding UCL to achieve the desired ARL$_{0}$.

\begin{algorithm}
\caption{Calculating Monitoring Statistic at time $t$}\label{alg:Meth}
\hrulefill \\
    \textbf{Input: }
        \begin{itemize}[label={},leftmargin=*]
            \item \texttt{Observed data: $\{\left(\bm{x}_{i}^{t},y_{i}^{t}\right)\}_{i=1}^n$}
            \item \texttt{Regression trees (or random forests):  $\mathcal{T}^{j}$, for $j=-m+1,\cdots,t-1$}
            \item \texttt{Empirical CDFs:  $\hat{F}_{e}^{j}$, for $j=-m+1,\cdots,t-1$}
        \end{itemize}
    \textbf{Output: } 
    \begin{itemize}[label={},leftmargin=*]
        \item \texttt{Monitoring statistic: $\xi^{t}$}
    \end{itemize}
    \textbf{Start algorithm}
    \begin{algorithmic}[1]
            \State \texttt{Obtain mean prediction using current data and all previous trees\newline 
                \hspace*{5em} $\hat{y}_{i}^{t}=\frac{1}{t-1+m} \sum_{j=-m+1}^{t-1} \mathcal{T}^{j} \left(\bm{x}_{i}^{t}\right)$} 
            \State \texttt{Calculate the residuals\newline 
                \hspace*{5em} $\hat{e}_{i}^{t}=y_{i}^{t}-\hat{y_{i}}^{t}$ for $i=1,\cdots, n$}
            \State \texttt{Calculate the empirical CDF $\hat{F}_{e}^{t}$ from $\hat{e}_{1}^{t},\ldots, \hat{e}_{n}^{t}$}
            \State \texttt{Calculate the monitoring statistic\newline 
                \hspace*{5em} $\xi^{t}=\max\left\{D_{n}\left(\hat{F}_{e}^{t},\hat{F}_{e}^{j}\right) \lvert j=-m+1,\cdots,t-1 \right\}$}
    \end{algorithmic}
    \textbf{End algorithm}
\end{algorithm}

{ 
\begin{algorithm}
    \caption{Simulating Run Length Using Bootstrap and Historical Data}\label{alg:BS}
    \hrulefill \\
    \textbf{Input: }
        \begin{itemize}[label={},leftmargin=*]
            \item \texttt{Historical data: $\left\{(\bm{x}_{i}^{j},y_{i}^{j})\right\}_{i=1}^n$, for $j=-m+1,\cdots,0$}
            \item \texttt{Regression trees (or random forests):  $\mathcal{T}^{j}$, for $j=-m+1,\cdots,0$}
            \item \texttt{Empirical CDFs:  $\hat{F}_{e}^{j}$, for $j=-m+1,\cdots,0$}
            \item \texttt{Control limit $\delta > 0$}
        \end{itemize}
    \textbf{Output: } 
    \begin{itemize}[label={},leftmargin=*]
        \item \texttt{Run Length: $T$}
    \end{itemize} 
    \textbf{Start algorithm}
    \begin{algorithmic}[1]
        \State \texttt{Set $\xi^{*} = 0$ and $t=0$}
        \State \texttt{Combine the $m$ historical data sets into a single one, $D = \bigcup_{j = -m+1}^0 \left\{(\bm{x}_{i}^{j},y_{i}^{j})\right\}_{i=1}^{n}$}
        \While{$\xi^{*} < \delta$}
            \State \texttt{Set $t=t+1$}
            \State \texttt{From the $m$~$\times$~$n$ data points in $D$, draw a sample of size $n$ with replacement and denote it as $\left\{(\bm{x}_{i}^{t},y_{i}^{t})\right\}_{i=1}^n$}
            \State \texttt{Find and save the regression tree (or random forest) $\mathcal{T}^t$ fitted based on $\left\{(\bm{x}_{i}^{t},y_{i}^{t})\right\}_{i=1}^n$}
            \State \texttt{Obtain mean prediction using current data and all previous trees\newline 
                \hspace*{5em} $\hat{y}_{i}^{t}=\frac{1}{t-1+m} \sum_{j=-m+1}^{t-1} \mathcal{T}^{j} \left(\bm{x}_{i}^{t}\right)$} 
            \State \texttt{Calculate the residuals\newline 
                \hspace*{5em} $\hat{e}_{i}^{t}=y_{i}^{t}-\hat{y_{i}}^{t}$ for $i=1,\cdots, n$}
            \State \texttt{Calculate and save the empirical CDF $\hat{F}_{e}^{t}$ from $\hat{e}_{1}^{t},\ldots, \hat{e}_{n}^{t}$}
            \State \texttt{Calculate the monitoring statistic\newline 
                \hspace*{5em} $\xi^{t}=\max\left\{D_{n}\left(\hat{F}_{e}^{t},\hat{F}_{e}^{j}\right) \lvert j=-m+1,\cdots,t-1 \right\}$}
            \If{{\color{black}$\xi^{t}\geq\delta$}}
                \State \texttt{Claim change-point} 
                \State \texttt{Record run length $T=t$}
            \EndIf
            \State $\xi^{*} = \xi^{t}$
        \EndWhile
    \end{algorithmic}
    \textbf{End algorithm}
\end{algorithm}
}

\section{Simulation Study}\label{sec:sim}

{\color{black} 

We conduct a simulation study to evaluate the performance of our proposed approach and compare it with the EWMA and SIM-based methods proposed by Li et al. \cite{Lietal} and Iguchi et al. \cite{Iguchi}, respectively.  Notably, Iguchi et al. previously compared their SIM-based method with Li et al.'s EWMA-based method and found that the former outperforms the latter in most cases.

For our simulations, we consider various data-generating mechanisms, signal-to-noise ratios (SNR), numbers of available historical profiles, and last in-control time-steps. We employ \autoref{alg:BS} to determine the UCL and compute the monitoring statistic using \autoref{alg:Meth}. The performance assessment is based on out-of-control Average Run Length (ARL$_1$) and False Alarm Rate (FAR), defined in Equations \ref{eq:ARL1} and \ref{eq:FArate}, respectively. In the subsequent two subsections, we present the details of our simulation setup and the results obtained.
}

\subsection{Simulation Setup}\label{subsec:simsetup}

We simulate data using the functional relationship presented in \autoref{eq:profile} in \autoref{Sec:Background}, assuming $p=3$ and $n=512$.
At time $t$, we sample our 3 explanatory variables independently from the uniform distribution with support $(0,1)$, that is $x^{t}_{i,k} \sim U(0,1)$ for $k=1,2,3$ and $i=1,\cdots,n$. The independent additive random noises in the profiles are sampled from a standard normal distribution with a mean of $0$ and variance of $1$, that is $\varepsilon^{t}_{i} \sim N(0,1)$. 
\begin{table}\caption{Factors for simulation study. The levels of the factors associated with the in-control and out-of-control profiles are provided in Equations \ref{eq:IClin}-\ref{eq:ICnlin} and \ref{eq:OCsin}-\ref{eq:OCndiff}, respectively.}\label{tab:SimFac} 
\centering
\begin{tabular}{ll}
\hline
\textbf{Signal-to-Noise Ratio} & SNR $\in\{3,5,7\}$                  \\
\textbf{Historic Profiles}     & $m\in\{20,40\}$                  \\
\textbf{Last in-control time-step}     & $\tau\in\{0,30\}$ \\
\textbf{In-control profiles}     & 2 profiles \\
\textbf{Out-of-control profiles}     & 3 profiles \\
\hline
\end{tabular}
\end{table}

To ensure a fair comparison with existing methods, we chose to use the same number of historic profiles $m$, time-step of last in-control profile $\tau$, and SNR as Iguchi et al. \cite{Iguchi}, which are shown in \autoref{tab:SimFac}. 
We also use the same in-control functions (Equations \ref{eq:IClin}-\ref{eq:ICnlin}) and out-of-control forcing functions 
(Equations \ref{eq:OCsin}-\ref{eq:OCndiff}) as Iguchi et al.:
\begin{align}
\textbf{Linear: } & \label{eq:IClin} f(\boldsymbol{x}) =  1 + 3x_{1} + 2x_{2} + x_{3} \\
\textbf{Non-linear: } & \label{eq:ICnlin} f(\boldsymbol{x}) =  (4/9) (3x_{1} + 2x_{2} + x_{3})^{2} \\\nonumber & \\
\textbf{Sinusoidal: } & \label{eq:OCsin} g(\boldsymbol{x}) = C \sin(2 \pi x_{1}x_{2} ) \\
\textbf{Non-differentiable: } & \label{eq:OCndiff} g(\boldsymbol{x}) = 25 |x_{1} - 0.5|e^{-x_{2}} I_{(x_{3} > 0.5)}(x) \\
\textbf{Localized change: } & \label{eq:OCHLoc} g(\boldsymbol{x}) = f(\boldsymbol{x}) + a I_{(\boldsymbol{x}\in \mathscr{R})} ,
\end{align} 
where $C=5$ in \autoref{eq:OCsin} for \autoref{eq:IClin} and $C=1$ in \autoref{eq:OCsin} for \autoref{eq:ICnlin} based on the work from Iguchi et. al \cite{Iguchi}.
We chose to add one additional out-of-control forcing function given in \autoref{eq:OCHLoc}, where $\mathscr{R}$ denotes a sphere of radius $r$ and centered at $(0.5,\ 0.5,\ 0.5)$.
This out-of-control forcing function is designed to test how well our method identifies a localized change in $f(\boldsymbol{x})$. Notice that this forcing function is equal to $f(\boldsymbol{x})$ if $\boldsymbol{x} \notin \mathscr{R}$, and equals to $f(\boldsymbol{x})+a$ if $\boldsymbol{x} \in \mathscr{R}$, where $a$ is a positive constant.


Equations \ref{eq:IClin}-\ref{eq:ICnlin} are used directly in the change-point detection framework given in \autoref{eq:icooc}, whereas the out-of-control function, $\phi(\boldsymbol{x})$, is a linear combination of $f(\boldsymbol{x})$ and $g(\boldsymbol{x})$ 
that is, 
\begin{equation}\label{eq:OCfunc}
    \phi(\boldsymbol{x}) = \lambda f(\boldsymbol{x}) + (1-\lambda) g(\boldsymbol{x})
\end{equation}
where $\lambda$ is a weight set to achieve a given SNR.

Following Iguchi et al., for Equations \ref{eq:OCsin} and \ref{eq:OCndiff}, 
SNR is defined as $\sigma^{2}_{\text{Signal}}/\sigma^{2}_{\text{noise}}$.
Recall, that we sample $\varepsilon_{i}^{t}\sim N(0,1)$, so $\sigma^{2}_{\text{noise}}=1$, leading to $\text{SNR}=\sigma^{2}_{\text{signal}}=\text{Var}\left( f(\boldsymbol{x})-\phi(\boldsymbol{x}) \right)$.  The values of the weights, $\lambda$, are obtained 
using grid search and by estimating the SNR using Monte Carlo simulation, as in \autoref{eq:SNR} with $S=10^6$.
The obtained results for the SNR values given in \autoref{tab:SimFac} are shown in \autoref{Tab:lamSNR}. 

\begin{equation}\label{eq:SNR}
     \widehat{SNR} = \frac{1}{S} \sum_{i=1}^{S} \left( f(\boldsymbol{x}_{i}) - \phi(\boldsymbol{x}_{i}) - \frac{1}{S} \sum_{j=1}^{S} \left ( f(\boldsymbol{x}_{j}) - \phi(\boldsymbol{x}_{j}) \right ) \right )^{2}
\end{equation}

\begin{table}
\caption{SNR-$\lambda$ and SNR-$a$ values for each in-control and out-of-control combination}
\centering
  \begin{tabular}{ | c | c | c | c | c | }
\hline
  \multirow{2}{*}{in-control Profile} & \multirow{2}{*}{out-of-control Profile} & \multicolumn{3}{c}{SNR-$\lambda$} \vline  \\
  \cline{3-5}
  & & 3 & 5 & 7 \\
  \hline
  \multirow{2}{*}{\autoref{eq:IClin}} & \autoref{eq:OCsin} with $C=5$ & 0.4568 & 0.2986 & 0.1699 \\
  \cline{2-5}
  & \autoref{eq:OCndiff} & 0.3945 & 0.2184 & 0.0752 \\
    \hline
    \multirow{2}{*}{\autoref{eq:ICnlin}} & \autoref{eq:OCsin} with $C=1$ & 0.4615 & 0.3048 & 0.1775 \\
  \cline{2-5}
  & \autoref{eq:OCndiff} & 0.5465 & 0.4146 & 0.3074 \\
    \hline\hline\hline
  \multirow{2}{*}{in-control Profile} & \multirow{2}{*}{out-of-control Profile} & \multicolumn{3}{c}{SNR-$a$ with $\lambda = 0$} \vline  \\
  \cline{3-5}
  & & 3 & 5 & 7 \\
  \hline
  {\autoref{eq:IClin}} & \multirow{2}{*}{\autoref{eq:OCHLoc} } & 
  \multirow{2}{*}{5.7735} & 
  \multirow{2}{*}{7.4535} & 
  \multirow{2}{*}{8.8191} \\
  \cline{1-1}
  {\autoref{eq:ICnlin}} &  &  & & \\
  \hline
\end{tabular}
  \label{Tab:lamSNR}
\end{table}
For the localized change forcing function, SNR can be computed analytically, {\color{black} SNR$=(1-\lambda)^2 a^2 v (1-v) $, where $v$ is the volume of the sphere $\mathscr{R}$. Notice that there are multiple combinations of $(\lambda, a, v)$ that lead to a given SNR. For this reason, we decided to fix the values of $\lambda$ and $v$ and make the SNR only a function of $a$. Since $v$ is a probability that results from integrating the uniform probability density function (i.e., the distribution of $\boldsymbol{x}$) over the sphere $\mathscr{R}$, we decided to fix it at $0.1$. $\lambda$ is set to $0$. Therefore, for a given SNR, the value of $a$ is determined by $a = (10/3) \sqrt{SNR}$.}
The values of $a$ for the SNR values of \autoref{tab:SimFac} are summarized in \autoref{Tab:lamSNR}.

\begin{figure}
  \centering
  \includegraphics[width=.4\textwidth]{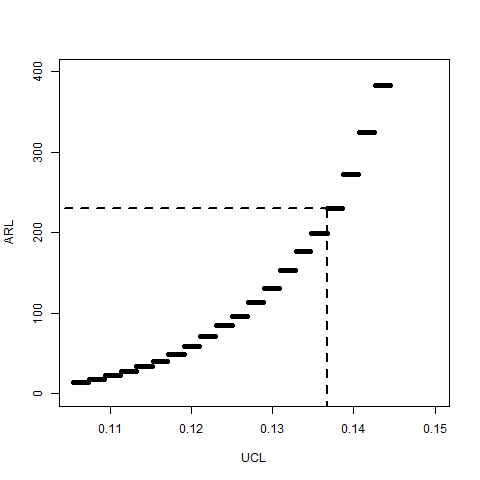}
  \caption{Typical pattern of ARL$_0$ as a function of the UCL}
  \label{fig:ARLUCLEx}
\end{figure}

Prior to the comparison, we must calibrate our model to find the UCL. The methods of Li et al. and Iguchi et al. provide continuous monitoring statistics and, as such, are calibrated for an in-control ARL$_{0}$ of 200. Our method, however, due to the discrete nature of the KS statistic in \autoref{eq:KSstatistics}, provides a discrete monitoring statistic. Consequently, it cannot achieve the desired ARL$_0$ of exactly 200. Therefore, using \autoref{alg:BS}, we opted to select the UCL that yields the smallest ARL$_0$ greater than 200. \autoref{fig:ARLUCLEx} illustrates how ARL$_0$ behaves as a function of the UCL for a specific simulated dataset. We observe a similar pattern across all simulated datasets.




For each $m$ and in-control function (Equations \ref{eq:IClin} and \ref{eq:ICnlin}), we simulate 100 sets of $m$ historic profiles. 
For each set of historic profiles, we run \autoref{alg:BS} 500 times to determine the UCL for both regression trees and random forests. 
Therein, we utilize the \texttt{R} packages \texttt{tree} \cite{RTree} and  \texttt{randomForests} \cite{RForest} to model the profiles based regression trees and random forests, respectively. The UCL's for all 100 sets of $m$ historic profiles are given in Tables \ref{tab:UCLsRT} and \ref{tab:UCLsrF} for regression trees and random forests, respectively. 

\begin{table}[t]\caption{Frequency of UCL's for all in-control combinations}\label{tab:UCLs}
\centering
\begin{subtable}[t]{0.4\textwidth}
\centering
\begin{tabular}[t]{lllll}
  \hline
\multicolumn{1}{c}{\multirow{2}{*}{UCL}} & \multicolumn{2}{c}{Linear}                          & \multicolumn{2}{c}{Non-Linear}                      \\ \cline{2-5} 
  \multicolumn{1}{c}{}                     & \multicolumn{1}{c}{m=20} & \multicolumn{1}{c}{m=40} & \multicolumn{1}{c}{m=20} & \multicolumn{1}{c}{m=40} \\ 
  \hline
  0.1348 &     &     &     &   1 \\
  0.1367 &  59 &  50 &  46 &  49 \\ 
  0.1387 &  21 &  22 &  29 &  19 \\ 
  0.1406 &   9 &  10 &   9 &  13 \\ 
  0.1426 &   4 &   7 &   6 &   5 \\ 
  0.1445 &   3 &   5 &   4 &   6 \\ 
  0.1465 &   1 &   4 &   3 &   1 \\ 
  0.1484 &   1 &   1 &   1 &   3 \\ 
  0.1504 &     &   1 &     &     \\
  0.1523 &     &     &     &   1 \\
  0.1543 &   1 &     &   2 &   1 \\ 
  0.1563 &   1 &     &     &     \\ 
  0.1582 &     &     &     &   1 \\
  0.1660 &     &     &     &     \\
   \hline
\end{tabular}\caption{Regression Trees}\label{tab:UCLsRT}
\end{subtable}
\begin{subtable}[t]{0.4\textwidth}
\begin{tabular}[t]{lllll}
  \hline
  \multicolumn{1}{c}{\multirow{2}{*}{UCL}} & \multicolumn{2}{c}{Linear}                          & \multicolumn{2}{c}{Non-Linear}                      \\ \cline{2-5} 
  \multicolumn{1}{c}{}                     & \multicolumn{1}{c}{m=20} & \multicolumn{1}{c}{m=40} & \multicolumn{1}{c}{m=20} & \multicolumn{1}{c}{m=40} \\ 
  \hline
  0.1348 & 1 &  & 2 & 2 \\
0.1367 & 66 & 65 & 86 & 85 \\
0.1387 & 17 & 14 & 9 & 8 \\
0.1406 & 5 & 10 & 2 & 3 \\
0.1426 & 4 & 1 &  &  \\
0.1445 & 3 & 8 &  &  \\
0.1465 & 1 &  & 1 &  \\
0.1484 & 2 &  &  &   \\
0.1504 &  & 1 &  & 1 \\
0.1523 &  &   &  &   \\
0.1543 &  & 1 &  & 1 \\
0.1563 &  &   &  &   \\ 
0.1582 &  &   &  &   \\
0.1660 & 1 &  &  &   \\
   \hline
\end{tabular}\caption{Random Forests}\label{tab:UCLsrF}
\end{subtable}
\end{table}


For each set of $m$ historic profiles (along with its corresponding UCL), SNR, in-control and out-of-control functions (Equations \ref{eq:IClin}-\ref{eq:OCHLoc}), and last in-control time-point $\tau$, we sequentially simulate a total of $\tau$ in-control profiles. Then, at time $t = \tau + 1$, we start simulating out-of-control profiles. At each time $t$, the monitoring statistic $\xi^{t}$ is computed using \autoref{eq:monitoring_statistic}, and we check if the process is out-of-control, that is, if $\xi^{t} \geq $UCL. The sequential simulation of profiles stops as soon as $\xi^{t}>$UCL with $t>\tau$. For each simulated sequence, the number of false alarms and its length (i.e., run length) are recorded.  \autoref{alg:Ph2} summarizes our Phase 2 simulations. Each set of conditions is evaluated through 50 runs of \autoref{alg:Ph2}. Finally, for each $m$, SNR, in-control and out-of-control functions (Equations \ref{eq:IClin}-\ref{eq:OCHLoc}), and $\tau$, we compute the false alarm rate (FAR) and average run length under the alternative hypothesis (ARL$_{1}$) using $N=5000$ trials (100 sets of historic profiles $\times$ 50 runs of \autoref{alg:Ph2}), based on \autoref{eq:FArate} and \autoref{eq:ARL1}, respectively.

\begin{algorithm} 
    \caption{Phase 2 Simulation}\label{alg:Ph2}
    \hrulefill \\
    \textbf{Input: }
        \begin{itemize}[label={},leftmargin=*]
            \item \texttt{Historical data: $\left\{(\bm{x}_{i}^{j},y_{i}^{j})\right\}_{i=1}^n$, for $j=-m+1,\cdots,0$}
            \item \texttt{Regression trees (or random forests):  $\mathcal{T}^{j}$, for $j=-m+1,\cdots,0$}
            \item \texttt{Empirical CDFs:  $\hat{F}_{e}^{j}$, for $j=-m+1,\cdots,0$}
            \item \texttt{Upper control limit: $\delta$}
            \item \texttt{Last in-control time-point: $\tau$}
            \item \texttt{Signal-to-noise ratio:$ $ SNR} 
            \item \texttt{In-control and out-of-control functions: $f$ and $\phi$}
        \end{itemize}
    \textbf{Output: } 
    \begin{itemize}[label={},leftmargin=*]
        \item \texttt{False Alarms:} $FA$
        \item \texttt{Run Length}: $T$
    \end{itemize}
    \textbf{Start algorithm}
    \begin{algorithmic}[1]
        \State \texttt{Set $\xi^{*} = 0$, $t=0$, and $FA=0$}
        \While{$\xi^{*} < \delta$}
            \State \texttt{Set $t=t+1$}
            \State \texttt{Sample $x_{i,k}^{t}\sim U(0,1)$ for $i=1,\cdots,512$ and $k = 1,2,3$}
            \State \texttt{Sample $\varepsilon_{i}^{t} \sim N(0,1)$ for $i=1,\cdots,512$}
            \If{$t\leq \tau$}
                \State \texttt{$y^{t}=f^{t}(\bm{x}_{i}^{t}) + \varepsilon_{i}^{t}$ for $i=1,\cdots,512$}
            \ElsIf{$t>\tau$}
                \State \texttt{$y_{i}^{t}=\phi^{t}(\bm{x}_{i}^{t}) + \varepsilon_{i}^{t}$ for $i=1,\cdots,512$}
            \EndIf
            \State \texttt{Find and save the regression tree (or random forest) $\mathcal{T}^t$ fitted based on $\left\{(\bm{x}_{i}^{t},y_{i}^{t})\right\}_{i=1}^n$}
            \State \texttt{Obtain mean prediction using current data and all previous trees\newline 
                \hspace*{5em} $\hat{y}_{i}^{t}=\frac{1}{t-1+m} \sum_{j=-m+1}^{t-1} \mathcal{T}^{j} \left(\bm{x}_{i}^{t}\right)$} 
            \State \texttt{Calculate the residuals\newline 
                \hspace*{5em} $\hat{e}_{i}^{t}=y_{i}^{t}-\hat{y_{i}}^{t}$ for $i=1,\cdots, n$}
            \State \texttt{Calculate and save the empirical CDF $\hat{F}_{e}^{t}$ from $\hat{e}_{1}^{t},\ldots, \hat{e}_{n}^{t}$}
            \State \texttt{Calculate the monitoring statistic\newline 
                \hspace*{5em} $\xi^{t}=\max\left\{D_{n}\left(\hat{F}_{e}^{t},\hat{F}_{e}^{j}\right) \lvert j=-m+1,\cdots,t-1 \right\}$}
            \If{$\xi^{t}\geq\delta$}
                \If{$t\leq \tau$}
                    \State \texttt{Record False Alarm, $FA=FA+1$}
                    \State Return to step 3
                \ElsIf{$t>\tau$}
                    \State \texttt{Claim Change-Point}
                    \State \texttt{Record run length $T=t$}
                \EndIf
            \EndIf
            \State $\xi^{*} = \xi^{t}$
        \EndWhile
    \end{algorithmic}
    \textbf{End algorithm}
\end{algorithm}

\subsection{Simulation Results}\label{subsec:simresults}

\begin{figure}
  \centering
  \includegraphics[trim=0.25cm 2.5cm 0.25cm 2.5cm, width=.95\textwidth]{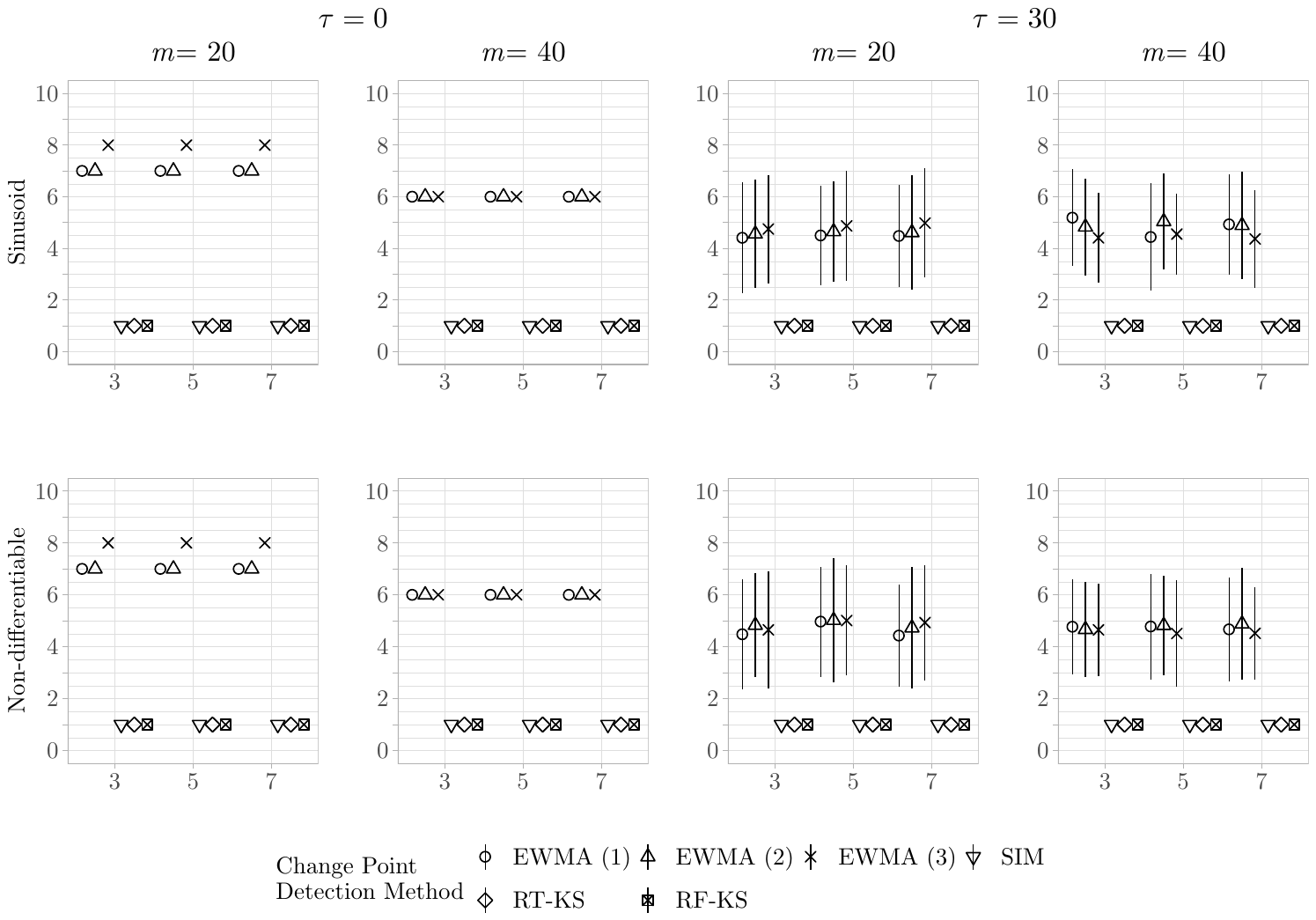}
  \caption{ARL$_{1}$ and errorbar plots for linear in-control function and for RT-KS, RF-KS, SIM, and EWMA methods. The figure displays ARL$_{1}$ results for sinusoidal (top row) and non-differentiable (bottom row) out-of-control forcing functions. Columns represent $\tau \in \{0,30\}$ and $m \in \{20,40\}$. Each panel shows the ARL$_{1}$ (y-axis) against the SNR (x-axis).}
  \label{fig:lin-ARL1}
\end{figure}
\begin{figure}
  \centering
  \includegraphics[trim=0.25cm 2.5cm 0.25cm 2.5cm,width=.95\textwidth]{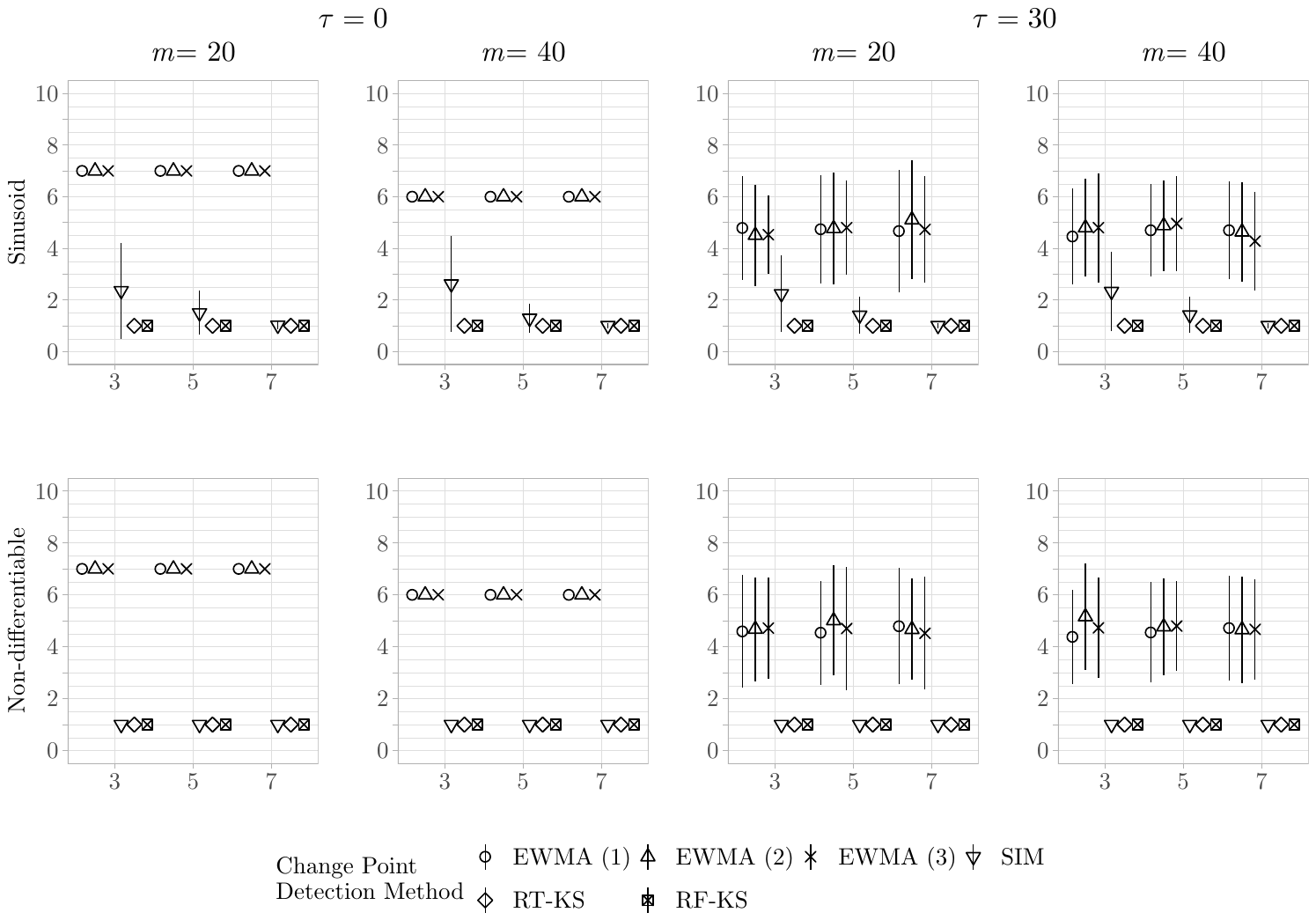}
  \caption{ARL$_{1}$ and errorbar plots for non-linear in-control function and for RT-KS, RF-KS, SIM, and EWMA methods. The figure displays ARL$_{1}$ results for sinusoidal (top row) and non-differentiable (bottom row) out-of-control forcing functions. Columns represent $\tau \in \{0,30\}$ and $m \in \{20,40\}$. Each panel shows the ARL$_{1}$ (y-axis) against the SNR (x-axis).}
  \label{fig:nlin-ARL1}
\end{figure}

We now analyze the results obtained from the simulations to assess the performance of our proposed methods using regression trees (RT-KS) and random forests (RF-KS) and compare them with the approach proposed by Li et al. \cite{Lietal} and Iguchi et al. \cite{Iguchi}. From now on, we refer to these two approaches simply as EWMA and SIM. For the sinusoidal (\autoref{eq:OCsin}) and non-differentiable (\autoref{eq:OCndiff}) out-of-control forcing functions, the results for ARL$_1$ are displayed in Figures \ref{fig:lin-ARL1} and \ref{fig:nlin-ARL1}.
We observe that our methods RT-KS and RF-KS achieved an ARL$_{1}$ approximately equal to the ideal value of $1$. They consistently outperformed EWMA. The ARL$_{1}$ for SIM was similar to RT-KS and RF-KS in most scenarios, but it struggled with the combination of non-linear in-control functions and sinusoidal out-of-control forcing functions. This struggle was particularly evident in the results for SNR$=\{3,5\}$.

\begin{figure}
  \centering
  \includegraphics[trim=0.25cm 2.5cm 0.25cm 2.5cm,width=.95\textwidth]{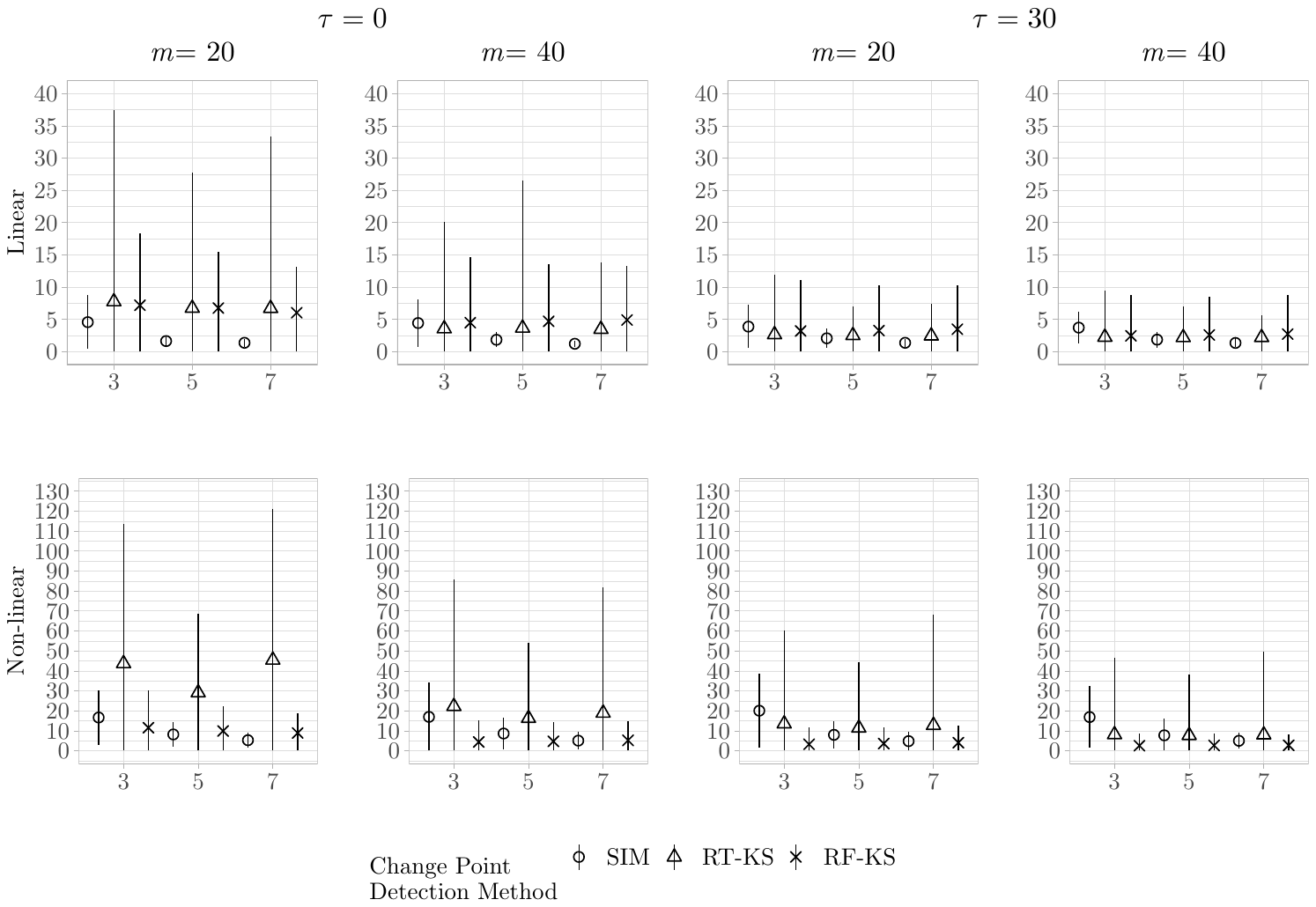}
  \caption{ARL$_{1}$ and errorbar plots for localized change out-of-control forcing function and for RT-KS, RF-KS, and  SIM methods. The figure displays ARL$_{1}$ results for linear in-control function (top row) and non-linear in-control function (bottom row) in-control functions. Columns represent $\tau \in \{0,30\}$ and $m \in \{20,40\}$. Each panel shows the ARL$_{1}$ (y-axis) against the SNR (x-axis).}
  \label{fig:HLoc-ARL1}
\end{figure}

Since RT-KS and RF-KS outperformed EWMA in every scenario, we only compared them to SIM using the localized change (\autoref{eq:OCHLoc}) out-of-control forcing function. The results for ARL$_{1}$ under this forcing function are shown in \autoref{fig:HLoc-ARL1}. Of the 24 scenarios (2 in-control functions $\times$ 3 SNR $\times$ 2 $\tau$ $\times$ 2 $m$), RF-KS outperformed RT-KS 14 times. The figure also shows that RF-KS had less variance (error bars) than RT-KS. When we compared our approaches to SIM, we found that out of the 24 scenarios, RT-KS outperformed SIM 6 times. Many of the results were very close. Of the 18 scenarios where SIM outperformed RT-KS, 3 of the $\text{ARL}_{1}$ values appeared to differ by only a fraction. Of the 6 where RT-KS outperformed SIM, 2 of them also differed in the $\text{ARL}_{1}$ value by only a fraction. For RF-KS, we outperformed SIM in 13 of the 24 scenarios.  

We can see there is a clear improvement from the EWMA models compared to the SIM method and our methods. However, the results from SIM and our methods are very close. In many cases, all three methods obtained an ARL$_{1}=1$. In the scenarios where all three of these methods did not identify the failure immediately (ARL$_{1}\ne1$), some of the results were too close to say that one method was clearly better. We decided to compare the ARL$_{1}$ values using the $t$-test to determine if the results are statistically better. Using the $t$-test with an $\alpha=0.05$, we compared SIM against both the RT-KS and the RF-KS methods, and we compared the RT-KS and RF-KS. In \autoref{tab:t-test}, we show the ARL$_{1}$ values for every scenario where SIM, RT-KS, and RF-KS did not all three have an ARL$_{1}=1$. The bold numbers in these tables indicate the lowest ARL$_{1}$ based on the results of the $t$-test. If more than one number is bold, then the ARL$_{1}$ is not statistically different. We can see that even though SIM had a higher ARL$_{1}$ for SNR$=7$ using the non-linear in-control function and sinusoidal out-of-control forcing function, it is not statistically different than the ARL$_{1}$ for our methods. We also see that RT-KS and RF-KS are not statistically different from each other for SNR$=3$ and $\tau=30$ using the linear in-control function and localized change out-of-control forcing function. There are two similar outcomes between SIM and RF-KS, both for the non-linear in-control function and localized change out-of-control forcing function. One when $\tau=0$, SNR=$7$, and $m=40$, while the other is when $\tau=30$, SNR=$7$, and $m=20$. 

\begin{table}\caption{A collection of $3\times1$ vectors for each combination of SNR, $m$, and $\tau$ containing the ARL$_{1}$ values for the non-linear in-control profile and sinusoidal and localized-change forcing functions. For each SNR, $m$, and $\tau$, three t-tests were performed (SIM versus RT-KS, SIM versus RF-KS, and RT-KS versus RF-KS), aiming to test whether there is any significant difference among the population ARL$_{1}$ values. A single bolded number within a vector indicates that the corresponding approach has the significantly lowest ARL$_{1}$. Two bolded values within a vector indicate that those approaches have the lowest ARL$_{1}$ values and the t-test comparing them did not detect any significant difference. When RF-KS and RT-KS have an ARL$_{1} = 1.00$, we only performed a single t-test to test if the population ARL$_{1}$ for SIM was equal to $1.00$. The ARL$_{1}$ for SIM was bolded if the hypothesis was not rejected. \color{red} }\label{tab:t-test}
\centering \vspace{5mm}
\begin{tabular}{cccccccccc}
                             &                                   &                       &        & \multicolumn{3}{c}{$m=20$}                                          & \multicolumn{3}{c}{$m=40$}                    \\ \cline{5-10} 
                             &                                   &                       &        & \multicolumn{3}{c}{SNR}                                             & \multicolumn{3}{c}{SNR}                       \\ \cline{5-10} 
In-control                   & Out-of-control                    & $\tau$                & Method & 3              & 5             & \multicolumn{1}{c|}{7}             & 3             & 5             & 7             \\ \hline
\multirow{6}{*}{Linear}      & \multirow{6}{*}{\begin{tabular}[c]{@{}c@{}}Localized\\      change\end{tabular}} & \multirow{3}{*}{$0$}  & SIM    & \textbf{4.58}  & \textbf{1.65} & \multicolumn{1}{c|}{\textbf{1.38}} & 4.44          & \textbf{1.85} & \textbf{1.23} \\
                             &                                   &                       & RT-KS  & 7.76           & 6.74          & \multicolumn{1}{c|}{6.72}          & \textbf{3.57} & 3.67          & 3.48          \\
                             &                                   &                       & RF-KS  & 7.2            & 6.77          & \multicolumn{1}{c|}{6.04}          & 4.5           & 4.7           & 4.92          \\ \cline{4-10} 
                             &                                   & \multirow{3}{*}{$30$} & SIM    & 3.89           & \textbf{2.08} & \multicolumn{1}{c|}{\textbf{1.38}} & 3.73          & \textbf{1.87} & \textbf{1.36} \\
                             &                                   &                       & RT-KS  & \textbf{2.65}  & 2.52          & \multicolumn{1}{c|}{2.46}          & \textbf{2.26} & 2.2           & 2.21          \\
                             &                                   &                       & RF-KS  & 3.21           & 3.28          & \multicolumn{1}{c|}{3.48}          & \textbf{2.47} & 2.6           & 2.74          \\ \cline{2-10} 
\multirow{12}{*}{Non-linear} & \multirow{6}{*}{\begin{tabular}[c]{@{}c@{}}Localized\\      change\end{tabular}} & \multirow{3}{*}{$0$}  & SIM    & 16.62          & \textbf{8.15} & \multicolumn{1}{c|}{\textbf{5.26}} & 16.94         & 8.61          & \textbf{5.04} \\
                             &                                   &                       & RT-KS  & 43.67          & 29.08         & \multicolumn{1}{c|}{45.43}         & 22.2          & 16.22         & 18.86         \\
                             &                                   &                       & RF-KS  & \textbf{11.53} & 9.87          & \multicolumn{1}{c|}{8.86}          & \textbf{4.35} & \textbf{4.68} & \textbf{5.18} \\ \cline{4-10} 
                             &                                   & \multirow{3}{*}{$30$} & SIM    & 20.03          & 7.95          & \multicolumn{1}{c|}{\textbf{4.79}} & 16.85         & 7.71          & 4.96          \\
                             &                                   &                       & RT-KS  & 13.47          & 11.37         & \multicolumn{1}{c|}{12.71}         & 8.08          & 7.69          & 7.97          \\
                             &                                   &                       & RF-KS  & \textbf{3.21}  & \textbf{3.6}  & \multicolumn{1}{c|}{\textbf{4.04}} & \textbf{2.51} & \textbf{2.64} & \textbf{2.7}  \\ \cline{3-10} 
                             & \multirow{6}{*}{Sinusoidal}       & \multirow{3}{*}{$0$}  & SIM    & 2.36           & 1.5           & \multicolumn{1}{c|}{\textbf{1.02}} & 2.24          & 1.41          & \textbf{1.01} \\
                             &                                   &                       & RT-KS  & \textbf{1.00}  & \textbf{1.00} & \multicolumn{1}{c|}{\textbf{1.00}} & \textbf{1.00} & \textbf{1.00} & \textbf{1.00} \\
                             &                                   &                       & RF-KS  & \textbf{1.00}  & \textbf{1.00} & \multicolumn{1}{c|}{\textbf{1.00}} & \textbf{1.00} & \textbf{1.00} & \textbf{1.00} \\ \cline{4-10} 
                             &                                   & \multirow{3}{*}{$30$} & SIM    & 2.62           & 1.29          & \multicolumn{1}{c|}{\textbf{1.01}} & 2.33          & 1.42          & \textbf{1.01} \\
                             &                                   &                       & RT-KS  & \textbf{1.00}  & \textbf{1.00} & \multicolumn{1}{c|}{\textbf{1.00}} & \textbf{1.00} & \textbf{1.00} & \textbf{1.00} \\
                             &                                   &                       & RF-KS  & \textbf{1.00}  & \textbf{1.00} & \multicolumn{1}{c|}{\textbf{1.00}} & \textbf{1.00} & \textbf{1.00} & \textbf{1.00} \\ \hline
\end{tabular}
\end{table}

\begin{figure}
  \centering
  \includegraphics[width=.8\textwidth] {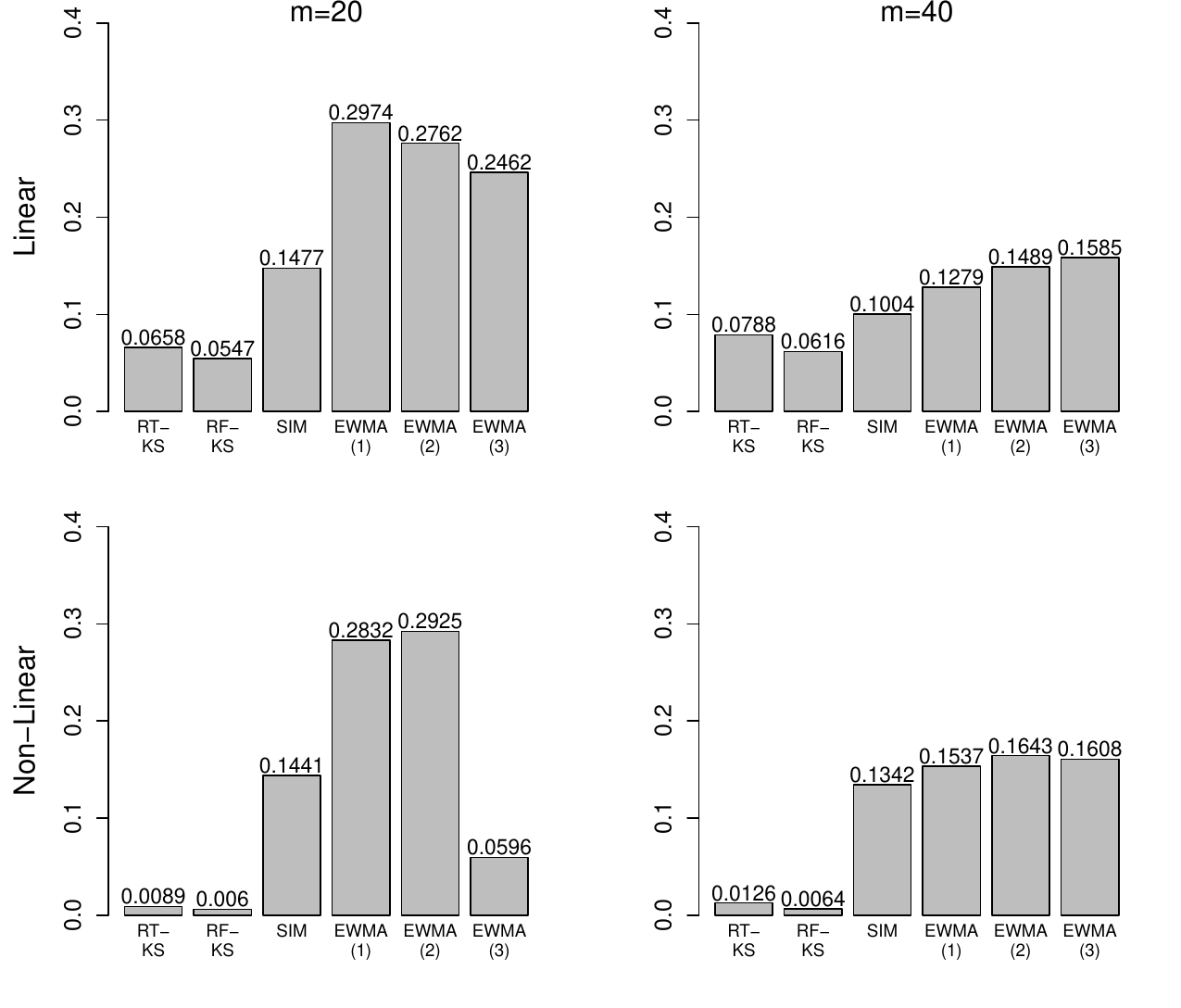}
  \caption{FAR plots for $\tau=30$. The figure displays FAR results for linear in-control function (top row) and non-linear in-control function (bottom row) in-control functions. Columns represent $m \in \{20,40\}$. Each panel shows the FAR (y-axis) against the methods (x-axis).}
  \label{fig:FAR}
\end{figure}

The other metric for comparison was the FAR given in \autoref{eq:FArate}. This performance metric only applied to the scenarios with $\tau=30$ since it had time to run while in-control. 
Since we have multiple scenarios from \autoref{tab:SimFac}, we combine the results from these simulations and group them based on the in-control function and number of historic profiles $m$. The results for the FAR are shown in \autoref{fig:FAR}. For all combinations of in-control functions and $m$, RT-KS and RF-KS had a much lower FAR than EWMA and SIM. When comparing RT-KS to RF-KS, we found that the latter had a slight edge for the FAR.

In more detail, when comparing our methods to those of competitors, we found that RT-KS had a $17\%-18\%$ lower ARL$_1$ and had a $73\%-81\%$ lower FAR than the EWMA method. For RF-KS, the ARL$_1$ rate was the same as for RT-KS, but the FAR was $79\%-85\%$ lower. RF-KS outperformed SIM more than RT-KS. On average, RT-KS had a $45\%$ higher ARL$_1$ than SIM, whereas RF-KS had a $27\%$ lower ARL$_1$. In terms of FAR, RT-KS had a $68\%$ lower rate, and RF-KS had a $76\%$ lower rate. On average, RT-KS had a $32\%$ higher ARL$_1$ than RF-KS and $29\%$ higher for FAR.

Since the SNR values listed in \autoref{tab:SimFac} (SNR $\geq 3$) led to our proposed methods achieving an ARL$_{1}$ approximately equal to the ideal value of $1$, we decided to further study this phenomenon. We used RT-KS and RF-KS, considering smaller SNR values, to determine where we are no longer able to achieve an ARL$_{1}$ of $1$. This was only done for the sinusoidal and non-differentiable out-of-control forcing functions, since the localized-change forcing function already does not result in an ARL$_{1}$ of $1$. We found that for both methods, they started to break down (with ARL$_{1}$ larger than $1$) around an SNR of $0.4$. Although the results for RT-KS and RF-KS were similar, RF-KS still outperformed RT-KS in terms of ARL$_{1}$.

\section{Discussion}
\label{sec:discuss}


We presented a novel nonparametric multivariate methodology for change-point detection for SPC utilizing regression trees, random forests, and the KS statistic. We have demonstrated that our proposed approach yields excellent results based on two performance metrics, ARL$_{1}$ and FAR, and that it compares favorably with other methods presented in the literature. When compared to each other, there is an apparent advantage in most scenarios for ARL$_{1}$ when using random forests versus rergession trees. However, this performance does come with a cost. The computational costs of using random forests in our methodology can be more than double that of regression trees as the monitoring time increases. If computational costs is not a consideration, then a practitioner may prefer to use random forests for better detection, otherwise we recommend the regression trees as they provide a low ARL$_{1}$ and lower computational cost.

One limitation of this methodology is its discrete monitoring statistic, resulting in an UCL leading to an ARL$_{0}$ greater than  the desired value of 200. However, this does not diminish the performance of our method when outperforming competitors, as a higher ARL$_{0}$ leads to a lower ability of out-of-control detection (i.e., lower ARL$_{1}$). Note that if the sample size increases, our KS-based monitoring statistic will have the ability to provide an ARL$_{0}$ closer to 200. This is because the empirical CDF, a step-wise function, will have smaller steps. Ongoing work focuses on creating a continuous version of the monitoring statistic capable of achieving any desired ARL$_{0}$ while still being as powerful as the proposed one.

\clearpage
\section*{Disclaimer}
The views expressed in this paper are those of the authors and do not reflect the official policy or position of the United States Air Force, Department of Defense, or the U.S. Government.
\section*{Acknowledgements}
We would like to thank \href{https://orcid.org/0000-0002-3029-5773}{\includegraphics[scale=0.06]{orcid.pdf}\hspace{1mm}Takayuki Iguchi} for all for his invaluable feedback.

This research was supported by the Office of the Secretary of Defense, Directorate of Operational Test and Evaluation, and the Test Resource Management Center under the Science of Test research program (FA8075-14-D-0019/ FA807518F1525). 

\bibliographystyle{ieeetr} 
\bibliography{references}  

\begin{thebibliography}{10}

\bibitem{SPC}
P.~Qiu, {\em Introduction to Statistical Process Control}.
\newblock Boca Raton, FL, USA: Taylor and Francis Group, 2014.

\bibitem{Nooro}
R.~Noorossana, A.~Saghaei, and A.~Amiri, {\em Statistical Analysis of Profile
  Monitoring}.
\newblock Wiley, 2011.

\bibitem{Wood-Sptiz-Monty}
W.~H. Woodall, D.~J. Spitzner, D.~C. Montgomery, and S.~Gupta, ``Using control
  charts to monitor process and product quality profiles,'' {\em Journal of
  Quality Technology}, vol.~36, pp.~309--320, 2004.

\bibitem{Woody_07}
W.~H. Woodall, ``Current research on profile monitoring,'' {\em Revista
  Producao}, vol.~17, pp.~420--425, 2007.

\bibitem{Woody-Monty}
W.~H. Woodall and D.~C. Montgomery, ``Some current directions in the theory and
  application of statistical process monitoring,'' {\em Journal of Quality
  Technology}, vol.~1, pp.~78--94, 2014.

\bibitem{gardner97}
M.~Gardner, J.-C. Lu, R.~Gyurcsik, J.~Wortman, B.~Hornung, H.~Heinisch,
  E.~Rying, S.~Rao, J.~Davis, P.~Mozumder, and et~al., ``Equipment fault
  detection using spatial signatures,'' {\em IEEE Transactions on Components,
  Packaging, and Manufacturing Technology: Part C}, vol.~20, no.~4,
  p.~295–304, 1997.

\bibitem{amiri_09}
A.~Amiri, W.~A. Jensen, and R.~B. Kazemzadeh, ``A case study on monitoring
  polynomial profiles in the automotive industry,'' {\em Quality and
  Reliability Engineering International}, vol.~26, no.~5, p.~509–520, 2009.

\bibitem{jin_99}
J.~Jin and J.~Shi, ``Feature-preserving data compression of stamping tonnage
  information using wavelets,'' {\em Technometrics}, vol.~41, no.~4,
  p.~327–339, 1999.

\bibitem{colosimo_08}
B.~M. Colosimo, Q.~Semeraro, and M.~Pacella, ``Statistical process control for
  geometric specifications: On the monitoring of roundness profiles,'' {\em
  Journal of Quality Technology}, vol.~40, no.~1, p.~1–18, 2008.

\bibitem{Florac_00}
W.~A. Florac, A.~D. Carleton, and J.~R. Barnard, ``Statistical process control:
  Analyzing space shuttle onboard software process,'' {\em IEEE Software},
  vol.~17, p.~97–106, 2000.

\bibitem{Cab1}
W.~N. Caballero, M.~Friend, and E.~Blasch, ``Adversarial machine learning and
  adversarial risk analysis in multi-source command and control,'' in {\em
  Signal Processing, Sensor/Information Fusion, and Target Recognition XXX}
  (I.~Kadar, E.~P. Blasch, and L.~L. Grewe, eds.), vol.~11756, pp.~98 -- 108,
  International Society for Optics and Photonics, SPIE, 2021.

\bibitem{Monty_09}
D.~C. Montgomery, {\em Introduction to statistical quality control}.
\newblock Wiley, 6~ed., 2009.

\bibitem{WilliWoodyBirch}
J.~D. Williams, W.~H. Woodall, and J.~B. Birch, ``Statistical monitoring of
  nonlinear product and process quality profiles,'' {\em Quality and
  Reliability Engineering International}, vol.~23, no.~8, pp.~925--941, 2007.

\bibitem{ChickPigSimpson}
E.~Chicken, J.~J. Pignatiello~Jr, and J.~R. Simpson, ``Statistical process
  monitoring of nonlinear profiles using wavelets,'' {\em Journal of Quality
  Technology}, vol.~41, no.~2, pp.~198--212, 2009.

\bibitem{Shing-Yada}
S.~I. Chang and S.~Yadama, ``Statistical process control for monitoring
  non-linear profiles using wavelet filtering and b-spline approximation,''
  {\em International Journal of Production Research}, vol.~48, no.~4,
  pp.~1049--1068, 2010.

\bibitem{NikooNooro}
M.~Nikoo and R.~Noorossana, ``Phase ii monitoring of nonlinear profile variance
  using wavelet,'' {\em Quality and Reliability Engineering International},
  vol.~29, no.~7, pp.~1081--1089, 2013.

\bibitem{McChicken}
K.~McGinnity, E.~Chicken, and J.~J. Pignatiello~Jr, ``Nonparametric changepoint
  estimation for sequential nonlinear profile monitoring,'' {\em Quality and
  Reliability Engineering International}, vol.~31, no.~1, pp.~57--73, 2015.

\bibitem{ChickHillPig}
E.~Chicken, R.~Hill, and J.~J. Pignatiello, ``Statistical functional process
  monitoring using clustered data,'' in {\em 67th Annual Conference and Expo of
  the Institute of Industrial Engineers}, vol.~1, p.~603–608, 2017 2017.

\bibitem{varbanov2019bayesian}
R.~Varbanov, E.~Chicken, A.~Linero, and Y.~Yang, ``A bayesian approach to
  sequential monitoring of nonlinear profiles using wavelets,'' {\em Quality
  and Reliability Engineering International}, vol.~35, no.~3, pp.~761--775,
  2019.

\bibitem{MagnificoGrass}
M.~Grasso, A.~Menafoglio, B.~M. Colosimo, and P.~Secchi, ``Using
  curve-registration information for profile monitoring,'' {\em Journal of
  Quality Technology}, vol.~48, no.~2, pp.~99--127, 2016.

\bibitem{ChuangHungYang}
S.-C. Chuang, Y.-C. Hung, W.-C. Tsai, and S.-F. Yang, ``A framework for
  nonparametric profile monitoring,'' {\em Computers \& Industrial
  Engineering}, vol.~64, no.~1, pp.~482--491, 2013.

\bibitem{HadSamHamid}
Z.~Hadidoust, Y.~Samimi, and H.~Shahriari, ``Monitoring and change-point
  estimation for spline-modeled non-linear profiles in phase ii,'' {\em Journal
  of Applied Statistics}, vol.~42, no.~12, pp.~2520--2530, 2015.

\bibitem{ZouQiuHawk}
C.~Zou, P.~Qiu, and D.~Hawkins, ``Nonparametric control chart for monitoring
  profiles using change point formulation and adaptive smoothing,'' {\em
  Statistica Sinica}, vol.~19, no.~3, 2009.

\bibitem{YangZouWang}
W.~Yang, C.~Zou, and Z.~Wang, ``Nonparametric profile monitoring using dynamic
  probability control limits,'' {\em Quality and Reliability Engineering
  International}, vol.~33, no.~5, pp.~1131--1142, 2017.

\bibitem{QieZou}
P.~Qiu and C.~Zou, ``Control chart for monitoring nonparametric profiles with
  arbitrary design,'' {\em Statistica Sinica}, vol.~20, no.~4, 2020.

\bibitem{HungTsaiYangChuang}
Y.-C. Hung, W.-C. Tsai, S.-F. Yang, S.-C. Chuang, and Y.-K. Tseng,
  ``Nonparametric profile monitoring in multi-dimensional data spaces,'' {\em
  Journal of Process Control}, vol.~22, no.~2, pp.~397--403, 2012.

\bibitem{Lietal}
C.-I. Li, J.-N. Pan, and C.-H. Liao, ``Monitoring nonlinear profile data using
  support vector regression method,'' {\em Quality and Reliability Engineering
  International}, vol.~35, p.~127–135, 2019.

\bibitem{Iguchi}
T.~Iguchi, A.~F. Barrientos, E.~Chicken, and D.~Sinha, ``Nonlinear profile
  monitoring with single index models,'' {\em Quality and Reliability
  Engineering International}, vol.~19, pp.~1--14, 2020.

\bibitem{breiman2001random}
L.~Breiman, ``Random forests,'' {\em Machine learning}, vol.~45, pp.~5--32,
  2001.

\bibitem{gareth2013introduction}
J.~Gareth, W.~Daniela, H.~Trevor, and T.~Robert, {\em An introduction to
  statistical learning: with applications in R}.
\newblock Spinger, 2013.

\bibitem{Kolmogorov}
A.~N. Kolmogorov, ``Sulla determinazione empirica di una legge di
  distribuzione,'' {\em Giornale dell'Instituto Italiano degli Attuari
  dell'Instituto Italiano degli Attuari}, vol.~4, p.~83–91, 1933.

\bibitem{Smirnov}
V.~I. Smirnov, ``Estimate of deviation between empirical distribution functions
  in two independent samples,'' {\em Bulletin Moscow University}, vol.~2,
  pp.~3--16, 1939.

\bibitem{Glivenko}
V.~I. Glivenko, ``Sulla determinazione empirica di una legge di
  distribuzione,'' {\em Giornale dell'Instituto Italiano degli Attuari
  dell'Instituto Italiano degli Attuari}, vol.~4, p.~92–99, 1933.

\bibitem{Cantelli}
F.~P. Cantelli, ``Sulla determinazione empirica di una legge di
  distribuzione,'' {\em Giornale dell'Instituto Italiano degli Attuari
  dell'Instituto Italiano degli Attuari}, vol.~4, p.~421–424, 1933.

\bibitem{Chernoff}
H.~Chernoff, ``A measure of asymptotic efficiency for tests of a hypothesis
  based on the sum of observations,'' {\em The Annals of Mathematical
  Statistics}, pp.~493--507, 1952.

\bibitem{RTree}
B.~Ripley, {\em Classification and Regression Trees}, 2021.
\newblock R package version >= 3.6.0.

\bibitem{RForest}
L.~Breiman, A.~Cutler, A.~Liaw, and M.~Wiener, {\em Classification and
  Regression based on a forest of trees using random inputs}, 2022.
\newblock R package version >= 4.1.0.

\end{thebibliography}










\end{document}